%% file: LatticePDFNLO.tex
\documentclass[twocolumn,aps,prl,superscriptaddress,nofootinbib,showpacs]{revtex4-1}
\usepackage{graphicx}
\usepackage{amsmath}
\usepackage{bm}
\usepackage{slashed}
\usepackage{epsfig}
\usepackage{amsfonts}
\usepackage{epstopdf}
\usepackage{hyperref}
\newcommand{\ben}{\begin{eqnarray}}
\newcommand{\een}{\end{eqnarray}}
\newcommand{\nnu}{\nonumber\\}
\newcommand{\bef}{\begin{figure}[!htp]}
\newcommand{\eef}{\end{figure}}

\newcommand{\bea}{\begin{eqnarray}}
\newcommand{\eea}{\end{eqnarray}}

\newcommand{\sect}[1]{\noindent{\it \textbf{#1} ---}}

\def\ba{\begin{linenomath*}\begin{equation}}
\def\ea{\end{equation}\end{linenomath*}}

\begin{document}
\title{Extracting Parton Distribution Functions from Lattice QCD Calculations}

\author{Yan-Qing Ma}
\email{yqma@pku.edu.cn}
\thanks{Previous address: Physics Department,
                 Brookhaven National Laboratory, Upton, NY 11973-5000, USA}
\affiliation{School of Physics and State Key Laboratory of Nuclear Physics and
Technology, Peking University, Beijing 100871, China}
\affiliation{Center for High Energy physics, Peking University, Beijing 100871, China}
\affiliation{Collaborative Innovation Center of Quantum Matter, Beijing 100871, China}
\author{Jian-Wei Qiu}
\email{jqiu@jlab.org}
\thanks{Previous address: Physics Department,
                 Brookhaven National Laboratory, Upton, NY 11973-5000; and
                 C.N.\ Yang Institute for Theoretical Physics
		and Department of Physics and Astronomy,
             	Stony Brook University, Stony Brook, NY 11794-3840, USA}
\affiliation{Theory Center, Jefferson Lab, 12000 Jefferson Avenue, Newport News, VA 23606, USA}

\date{\today}

\begin{abstract}
Parton distribution functions (PDFs) are nonperturbative quantities describing the relation between a hadron and  quarks and gluons within it.  We propose to extract PDFs from QCD global analysis of  ``data" generated by lattice QCD calculations of good ``lattice cross sections", which are basically single-hadron matrix elements that are lattice QCD calculable and perturbative QCD factorizable into the PDFs. To demonstrate the existence of good ``lattice cross sections", we take quasi-quark distribution introduced by Ji \cite{Ji:2013dva} as a case study to show that it could be factorized into the PDFs to all orders in perturbation theory if it can be multiplicatively renormalized.  We calculate the factorized coefficients at the next-to-leading order in $\alpha_s$.
\end{abstract}

\maketitle
\allowdisplaybreaks

\sect{Introduction}
Quantum Chromodynamics (QCD) is believed to be the fundamental theory for strong interactions, and is responsible for holding the quarks and gluons (or in general, partons) together to form nucleons and nuclei, the core of all visible matter in the universe. However, owing to the color confinement of QCD dynamics, no quarks and gluons have ever been observed in isolation by any detector in high energy scattering experiments.  It is the QCD factorization \cite{Collins:1989gx} that enables us to connect the QCD dynamics of quarks and gluons to physically measured hard scattering cross sections of identified hadrons.  Parton distribution functions (PDFs), $f_{i}(x,\mu)$, defined as the probability distributions to find a quark or a gluon ($i=q,\bar{q},g$) in a hadron carrying the hadron's momentum fraction between $x$ and $x+dx$, probed at the factorization scale $\mu$, are nonperturbative quantities to link the short-distance scattering between quarks and gluons to the confinement-sensitive physics of the colliding hadron(s).

Enormous theoretical and experimental efforts have been devoted to the extraction of PDFs by QCD global analysis of all existing high energy scattering data in the framework of QCD factorization \cite{Gao:2013xoa,Martin:2009iq,Ball:2011mu,Alekhin:2013nda}.  Although PDFs are well-defined matrix elements of quark and/or gluon fields in QCD, it is difficult, if not impossible, to calculate them directly in QCD due to their nonperturbative nature.  Since the operators defining PDFs depend on Minkowski time, even lattice QCD (LQCD) with an Euclidean time has only managed to calculate a few moments of PDFs \cite{Dolgov:2002zm,Gockeler:2004wp}.  Recently, Ji \cite{Ji:2013dva} introduced a set of quasi-PDFs,
and suggested that quasi-PDFs of hadron momentum $P_z$ become corresponding PDFs when $P_z$ is boosted to infinity.  However, since the hadron momentum in LQCD calculation is effectively bounded by lattice spacing, the $P_z\to\infty$ limit is hard to take in practice.  The connection between the quasi-PDFs and PDFs is further complicated by the fact that quasi-PDFs are not defined by the twist-2 operators, and have power ultra-violet (UV) divergences, while PDFs have only logarithmic UV divergences.

In this paper, we propose a rigorous program to {\it extract} PDFs from {\it good} ``time-independent''  and LQCD calculable hadronic matrix elements, defined with equal-time operators, that can be factorized into the desired PDFs in the same sprit of QCD factorization of experimental cross sections.  We refer these {\it good} hadronic matrix elements as ``lattice cross sections'' (LCSs).  PDFs can be extracted from QCD global analysis of LQCD generated data of LCSs, similar to what we have done to extract PDFs from experimental data.  We can continue improving our knowledge of PDFs by identifying more good LCSs and performing factorization of these LCSs with better calculated perturbative coefficient functions.  With the current computing power and lattice size, this program provides a unique opportunity to explore PDFs in the large $x$ region, where PDFs extracted from fitting experimental data still have huge uncertainties \cite{Gao:2013xoa,Martin:2009iq,Ball:2011mu,Alekhin:2013nda}.

The key for this program to work is the existence of good LCSs, and the robustness of their {\it calculability} in LQCD and {\it factorizability} in perturbative QCD (pQCD). In this paper, we will concentrate on the quasi-quark distribution as a case study to demonstrate that it could be a good LCS for extracting PDFs, and leave the effort to identify and construct more good LCSs to a future publication \cite{Ma:2017pxb}.  Since quasi-quark distribution is LQCD calculable \cite{Lin:2014zya}, we will show below that quasi-quark distribution could be factorized into PDFs to all orders in QCD perturbation theory with perturbatively calculable coefficient functions.  In addition, we also derive factorized coefficient functions at next-to-leading order (NLO) in $\alpha_s$ to confirm that they are infrared (IR) safe.

Our proposed program to extract PDFs from LQCD calculable LCSs could be extended for extracting other parton distributions and correlation functions by constructing new LCSs, including matrix elements made of two states of different momenta.  In addition, this program could provide new opportunities to explore partonic structure of hadrons, such as free neutron or various mesons, that we have difficulties to do scattering experiments with.
With this program and LQCD generated data on LCSs, along with experimentally measured data on hadronic cross sections, we could develop a comprehensive ``view'' of quark-gluon structure of hadrons.

\sect{Lattice cross section (LCS)}
We define a coordinate-space inclusive LCS as a single-hadron matrix element of a composite nonlocal operator ${\cal O}(\xi)$ made of quark/gluon fields or currents of quark/gluon fields, $\sigma(P\cdot\xi,\xi^2, \tilde{\mu}^2)\equiv \langle h(P)|T\{{\cal O}(\xi)\}|h(P)\rangle$, where $T$ stands for time-ordering, $\tilde{\mu}$ is the renormalization scale, $P$ is the momentum of the hadron $h$, and $\xi$ is the largest separation between fields or currents. In addition, to simplify our discussion, we have assumed $\xi^2$ to be small enough so that $\xi^2 P^2$ is negligible.  The momentum $P$ and the separation $\xi$ define the kinematics of the LCS, with $P\cdot\xi/\xi^2$  as the center of mass ``collision energy'' and $1/\xi^2$ defining the ``hard scale''.
In order to ensure that we can extract PDFs, a good LCS should have the following properties:
\begin{itemize}
\item
is calculable in LQCD with an Euclidean time,
\item
has a well-defined continuum limit as the lattice spacing $a\to 0$, and
\item
has the same and factorizable logarithmic collinear (CO) divergences as PDFs.
\end{itemize}
The first property could be satisfied by setting $\xi_0=0$ in $\sigma( P\cdot\xi, \xi^2, \tilde{\mu}^2)$, the second property is closely connected to the renormalizability of the operator, ${\cal O}(\xi)$, and it is the last property that enables us to extract PDFs from LQCD calculations. Our strategy to extract PDFs from good LCSs could be summarized by the following schematic plot,
\begin{eqnarray}
\begin{split}
\overline{\sigma}_{\text{E}}^\text{Lat}(P\cdot\xi,\xi^2,1/a^2)
\overset{\cal Z}\longleftrightarrow
{\sigma}_{\text{E}}&(P\cdot\xi,\xi^2,\,\tilde{\mu}^2)
\label{eq:matching}\\
&\Updownarrow\\
{\sigma}_{\text{M}}(P\cdot\xi,\xi^2,\,&\tilde{\mu}^2)
\overset{\cal C}\longleftrightarrow
f_{i}(x,\mu^2)\, ,
\end{split}
\label{eq:matchings}
\end{eqnarray}
where  $\overline{\sigma}_{\text{E}}^\text{Lat}(P\cdot\xi,\xi^2,1/a^2)$ is the discretized and LQCD calculated version of LCS with $\sigma_{\text{E}}(P\cdot\xi,\xi^2,\tilde{\mu}^2)$ as its renormalized continuum limit, and ${\sigma}_{\text{M}}(P\cdot\xi,\xi^2,\tilde{\mu}^2)$ is the Minkowski space version of ${\sigma}_{\text{E}}(P\cdot\xi,\xi^2,\tilde{\mu}^2)$ as indicated by its subscript ``M''.  The two lines in Eq.~(\ref{eq:matchings}) effectively represent the two key components of our proposed program:~{\it calculability} and {\it factorizability}, respectively.  For the calculability, we need to generate the ``data'' of $\overline{\sigma}_\text{E}^\text{Lat}(P\cdot\xi,\xi^2,1/a^2)$ from LQCD calculation for various good LCSs, correcting them for the continuum limit with a proper renormalization at the scale $\tilde{\mu}$ and matching coefficient functions ${\cal Z}$'s.  For the factorizability, we extract PDFs by performing QCD global analysis of the ``data'' with the following factorization formalism and pQCD calculated matching coefficients ${\cal C}$'s,
\begin{eqnarray}
{\sigma}(P\cdot\xi,\xi^2,\tilde{\mu}^2)
&\approx &
\sum_{i=q,\bar{q},g} \int_0^1 \frac{dx}{x}\,f_{i}(x,\mu^2)\,
\nonumber\\
& & {\hskip -0.1in}
\times\,
{\cal C}_i(x P\cdot\xi,\xi^2,  \tilde{\mu}^2,\mu^2) + {O}\left({\xi^2}\right),~~
\label{eq:factorization}
\end{eqnarray}
where we neglected the subscript ``M'' for the simplicity and $\mu$ is the factorization scale. 
This procedure is effectively the same as the traditional QCD global analysis to extract PDFs from high energy scattering data.

Our program to extract PDFs from the LQCD calculated data of LCSs sketched in Eq.~(\ref{eq:matchings}) should also work for LCSs in momentum-space,  if the Fourier Transformation (F.T.) from the coordinate-space is well behaved.  We define a momentum-space LCS, $\widetilde{\sigma}(\tilde{x},Q^2,\tilde{\mu}^2)$, in terms of F.T. of  $\sigma( P\cdot\xi, \xi^2, \tilde{\mu}^2)$ over the $\xi$ with a dimensionless variable $\tilde{x}$ as a Fourier conjugate of the variable $P\cdot\xi$, and $Q^2\sim \tilde{x}^2[\vec{P}|^2$ as the momentum-space hard scale.

In order to identify good LCSs, we need to demonstrate both the {\it calculability} and {\it factorizability} for  potential single hadron matrix elements, as specified in Eq.~(\ref{eq:matchings}). In the following, as a complete case study, we will show that the quasi-quark distribution could be a good LCS.  Since quasi-quark distribution is calculable in LQCD and its renormalizability has been well studied in Refs.~\cite{Ji:2017oey,Ishikawa:2017faj,Green:2017xeu}, we will show below that quasi-quark distribution can be factorized to PDFs with perturbatively calculable coefficients to all orders in QCD perturbation theory.

\sect{Quasi-PDFs}
The quasi-quark distribution, introduced by Ji \cite{Ji:2013dva}, is a special case of LCS in momentum-space with the operator ${\cal O}(\xi) = \overline{\psi}(\xi) \gamma\cdot\xi\, \Phi_{\xi}^{(f)}(\{\xi,0\}) \psi(0)$ where $\xi=(\xi_0,\xi_\perp,\xi_z)$,
\begin{eqnarray}
\label{eq:ftq}
\tilde{f}_{q/h}(\tilde{x},P_z,\tilde{\mu}^2)
&\equiv &
\int \frac{d\xi_z}{\pi} e^{-i \tilde{x}P_z \xi_z}\langle h(P)| \overline{\psi}(\xi_z)\,\frac{\gamma_z}{2}
\nonumber\\
&\ & \times
\Phi_{n_z}^{(f)}(\{\xi_z,0\})\, \psi(0) | h(P) \rangle\, ,
\end{eqnarray}
where $\xi_0=\xi_\perp=0$, the gauge link $\Phi_{n_z}^{(f)}(\{\xi_z,0\}) = \text{exp}[-ig\int_0^{\xi_z} d\eta_z\, A^{(f)}_z(\eta_z)]$ with the superscript ``$(f)$'' representing the fundamental representation of QCD's SU(3) color, and $n_z^\mu = (0,0_\perp, 1)$ with $n_z^2 = -1$ and $v\cdot n_z = -v_z$ for any vector $v^\mu$. After $\xi_z$ is integrated out, the hard scale for $\tilde{f}_{q/h}(\tilde{x},P_z,\tilde{\mu}^2)$ is $\tilde{x}P_z$, which is conjugate to $\xi_z$.  Similarly, quasi-gluon distribution is defined accordingly \cite{Ji:2013dva}.   As defined, the quasi-PDFs are gauge invariant and could be calculated in lattice QCD \cite{Lin:2014zya}.  But, unlike the PDFs, these quasi-PDFs are not boost invariant, and therefore, depend on the hadron momentum $P_z$. Their ``momentum fraction'', $\tilde{x}=k_z/P_z \in(-\infty,\infty)$ is not bounded by $P_z$, and they do not conserve the total ``parton'' momentum,
\begin{eqnarray}
\widetilde{\cal M}
\equiv
\sum_{i=q,\bar{q},g}
\int_{0}^\infty {\hskip -0.05in}
d\tilde{x}\, \tilde{x}\, \tilde{f}_{i/h}(\tilde{x},P_z,\tilde{\mu}^2)
\neq \text{constant}\, .
\label{eq:momentum}
\end{eqnarray}
Like the PDFs, the operators defining the quasi-PDFs have UV divergences and require UV renormalization.

\sect{Factorization of quasi-quark distribution}
We will show that quasi-quark distribution has logarithmic CO divergences, which could be systematically factorized into PDFs with IR safe coefficients, if its UV divergences could be renormalized multiplicatively \cite{Ji:2017oey,Ishikawa:2017faj}.

\begin{figure}[h]
\begin{minipage}[c]{0.6in}
\psfig{file=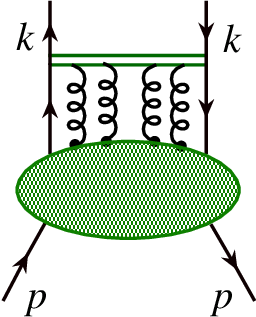,width=0.6in}
\end{minipage}
= \
\begin{minipage}[c]{0.5in}
\psfig{file=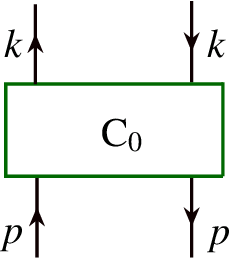,width=0.5in}
\end{minipage}
\ + \
\begin{minipage}[c]{0.5in}
\psfig{file=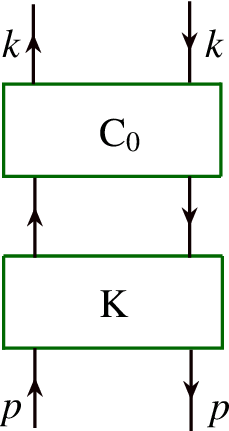,width=0.5in}
\end{minipage}
\ + \
\begin{minipage}[c]{0.5in}
\psfig{file=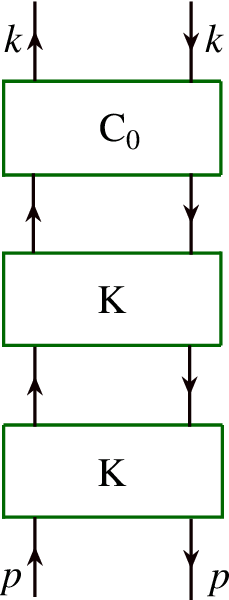,width=0.5in}
\end{minipage}
\ + ...
\caption{\label{fig:quasi-quark}
Ladder expansion of the quasi-quark distribution.}
\end{figure}

Like normal quark distribution, quasi-quark distribution can be represented by the forward scattering Feynman diagram, as shown in the left of Fig.~\ref{fig:quasi-quark}, with the active quark of momentum $k$ contracted with the ``cut-vertex'', $\gamma_z/(2P_z) \delta(\tilde{x}-k_z/P_z)$.  The gauge link in Eq.~(\ref{eq:ftq}) is represented by the double lines in Fig.~\ref{fig:quasi-quark}. Following effectively the same arguments used in Ref.~\cite{Collins:1981uw}, it is straightforward to show that the quasi-quark distribution of an asymptotic {\it parton state} is free of IR divergence.

In the following, to simplify our discussion, we show the proof of factorization for flavor non-singlet quasi-quark distribution, and the proof for flavor singlet case can be obtained similarly.
We find that if we expand the quasi-quark distribution of an asymptotic parton state of momentum $p$ in powers of $1/(\tilde{x}P_z)^2$, when it is small, the leading power contributions in the light-cone $n\cdot A=0$ gauge with the light-cone vector $n^\mu=(n^+,n^-,n_\perp)=(0,1,0_\perp)$ can be represented by a sum of ladder diagrams, as shown in Fig.~\ref{fig:quasi-quark}, where $C_0$ and $K$ are two-particle irreducible (2PI) kernels \cite{Ellis:1978ty}.
\begin{figure}[h]
\begin{minipage}[c]{0.8in}
\psfig{file=./figures/kernel-C0,width=0.6in}
\end{minipage}
\ = \
\begin{minipage}[c]{0.8in}
\psfig{file=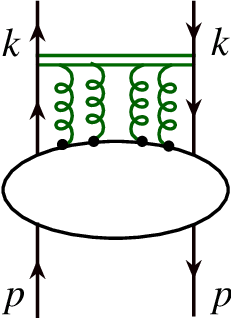,width=0.6in}
\end{minipage}
\caption{\label{fig:quasi-quark-q0}
Process-dependent 2PI kernel with the gauge link along the direction $n_z^\mu$ for quasi-PDFs, and multiplicative UV renormalization factors included.}
\end{figure}
By definition, $K$ includes the two quark propagators connecting to the kernel above. The dependence on the operator definitions of quasi-quark distribution is included in $C_0$, as shown in Fig.~\ref{fig:quasi-quark-q0}. The multiplicative renormalizability of quasi-quark distribution proved in Refs. \cite{Ji:2017oey,Ishikawa:2017faj,Green:2017xeu} implies that the renormalization of quasi-quark distribution is local and fully included in $C_0$, the renormalized version of which is denoted as $C_{\text{ren}}$.  In general, the renormalized 2PI kernels with fixed external momenta are finite in a physical gauge, such as the light-cone gauge \cite{Ellis:1978ty}.

\begin{figure}[h]
\begin{minipage}[c]{0.8in}
\psfig{file=./figures/kernel-CK0,width=0.5in}
\end{minipage}
\ $\Rightarrow$ \
\begin{minipage}[c]{0.7in}
\psfig{file=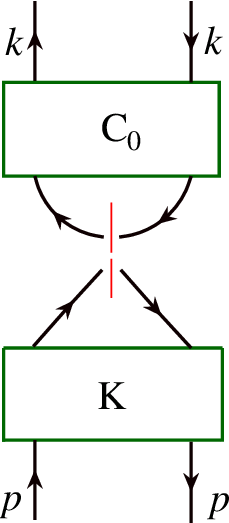,width=0.5in}
\end{minipage}
\begin{minipage}[c]{1.2in}
\leftline{$\leftarrow \frac{1}{2}\gamma\cdot p\, dx_i$}
\break
\leftline{$\leftarrow \frac{\gamma\cdot n}{2p\cdot n}\,\delta
{\hskip -0.02in}\left(x_i-\frac{k_i\cdot n}{p\cdot n}\right)$}
\end{minipage}
\caption{\label{fig:spin-2pi}
Spinor decomposition between two 2PI kernels.}
\end{figure}

To identify the leading power CO divergences, the spinor trace between two neighboring kernels can be approximated by the decomposition in Fig.~\ref{fig:spin-2pi}. The integration of the loop momentum $k_i$ between two neighboring 2PI kernels can be written as $\int d^4k_i = \int dx_i\int d^4k_i\, \delta(x_i-k_i\cdot n/p\cdot n)$, and be reduced to an one-dimensional integration $\int dx_i$, if we approximate the momentum $k_i$ entering the top kernel as $k_i\approx x_ip\cdot n$.  Since the $\gamma\cdot n/2p\cdot n\,\delta(x_i-k_i\cdot n/p\cdot n)$ is the cut-vertex defining the normal quark distribution, the phase space integration, $\int d^4k_i$ over $K$, with its two lower quark lines contracted with $\gamma\cdot p/2$, gives the well-known perturbative logarithmic UV and CO divergences of the quark distribution function \cite{Collins:1981uw}.  A standard UV renormalization for PDFs should remove all logarithmic UV divergences associated with this phase space integration of $K$.
Generally, we introduce a projection operator, $\widehat{\cal P}$ with $\widehat{\cal P}{W} \equiv \int d^4k_i/(2\pi)^4\delta(x_i-k_i\cdot n/p\cdot n)\, \text{Tr}[\gamma\cdot n\, W\, \gamma\cdot p]/(4p\cdot n)  + \text{UVCT}$ with the UV counter-term (UVCT), to pick up the CO divergence of any bottom part of diagram ${W}$ with its logarithmic UV divergences removed. Then $(1-\widehat{\cal P}){W}$ is free of CO divergence.

Symbolically, the renormalized quasi-quark distribution of a parton state  in Fig.~\ref{fig:quasi-quark}  can be expressed as $\tilde{f}_{q/h}^{\text{ren}}= {C}_\text{ren} \sum_{i=0}^\infty {K}^i$.  To factorize all of its perturbative CO divergences into PDFs of the same parton state, we decompose the last $K$ by $\widehat{\cal P}{K}$ and $(1-\widehat{\cal P}) {K}$,
\begin{align}\label{eq:fac-co0}
\tilde{f}_{q/h}^{\text{ren}}
 &=
 {C}_\text{ren}
{\hskip -0.03in} \bigg[
1 + \sum_{i=0}^{\infty} {K}^{i} (1-\widehat{\cal P}) {K}\bigg]
+ \tilde{f}_{q/h}^{\text{ren}}\, \widehat{\cal P}\, {K}\, .
\end{align}
Similarly, we can then decompose $K (1-\widehat{\cal P}) {K}$. Following this procedure repeatedly, we eventually arrive at
\begin{eqnarray}\label{eq:fac-co}
\tilde{f}_{q/h}^{\text{ren}} \bigg\{ 1-\, \widehat{\cal P}\, {K}
{\hskip -0.03in}
\sum_{i=0}^{\infty} \big[(1-\widehat{\cal P}) {K}\big]^i\bigg\}
{\hskip -0.1in} &=& {\hskip -0.08in}
 {C}_\text{ren}
{\hskip -0.03in}
\sum_{i=0}^{\infty} \big[(1-\widehat{\cal P}) {K}\big]^i {\hskip -0.08in} .~~~~
\end{eqnarray}
Now it is ready to obtain a factorized form by dividing both sides by the CO divergent factor  \cite{Ellis:1978ty}, which gives
\begin{equation}
\tilde{f}_{q/h}^{\text{ren}} = \left[
{C}_\text{ren} \frac{1}{1-(1-\widehat{\cal P}){K}}\right]
\left[
\frac{1}{1-\widehat{\cal P}{K}}
\right]\,,
\label{eq:fac}
\end{equation}
where all CO divergences of the renormalized quasi-quark distribution are now factorized into the second term, which is equal to the perturbative contribution to the normal quark distribution.
The derivation of  Eq.~(\ref{eq:fac}) could be easily extended to the CO factorization of quasi-gluon distribution if it can be ``renormalized''.  That is, multiplicatively {\it renormalized} quasi-PDFs share the same CO divergence as that of PDFs, since all quasi-PDFs dependence are included in $C_\text{ren}$, and could be factorized into PDFs as in Eq.~(\ref{eq:factorization}) plus the power corrections,
\begin{eqnarray}
\tilde{f}_{q/h}^{\text{ren}} &= &
 f_{i/h}\otimes\,
{\cal C}_{q/i}+ O((\tilde{x}P_z)^{-2})\,.
\label{eq:facQuasi}
\end{eqnarray}

\sect{The coefficient functions at NLO}
We calculate the coefficient function in Eq.~(\ref{eq:factorization}) for quasi-quark distribution at NLO to explicitly verify that it is free of IR and CO divergences.

By expanding both sides of Eq.~\eqref{eq:facQuasi} to order $\alpha_s^0$ and using the normalization $\tilde{f}^{(0)}_{q/q}(\tilde{x}) = \delta(1-\tilde{x})$ and $f^{(0)}_{q/q}(x) = \delta(1-x)$, where we neglected the superscript ``ren'' for simplicity, we obtain ${\cal C}_{q/q}^{(0)}({t})= \delta(1-{t})$ with ${t}=\tilde{x}/x$.
\begin{figure}[h]
\psfig{file=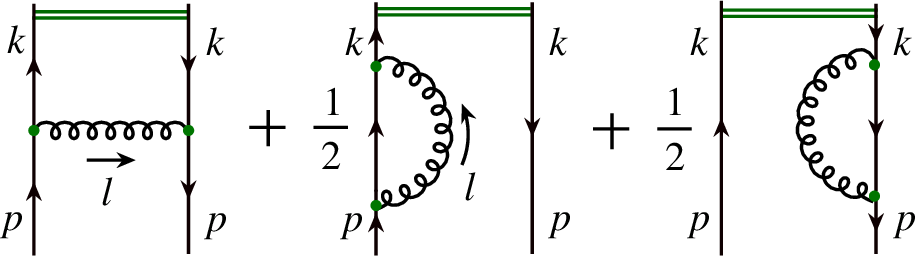,width=2.4in}
\caption{\label{fig:quasi-quark-lo}
Next-to-leading order diagrams contribute to the quasi-quark distribution of a quark.}
\end{figure}
By expanding both sides of the equation~(\ref{eq:facQuasi}) to order $\alpha_s$, and keeping only the flavor nonsinglet contribution, we obtain
\begin{eqnarray}
\tilde{f}_{q/q}^{(1)}(\tilde{x})
=
f^{(0)}_{q/q}(x) \otimes
{\cal C}_{q/q}^{(1)}(\tilde{x}/x)
+
f^{(1)}_{q/q}(x) \otimes
{\cal C}_{q/q}^{(0)}(\tilde{x}/x)\, ,
\label{eq:1st}
\end{eqnarray}
where $\otimes$ represents the convolution over $x$ in Eq.~(\ref{eq:factorization}).
Using the LO results above, we obtain
\begin{eqnarray}
{\cal C}_{q/q}^{(1)}({t},P_z,\tilde{\mu}^2,\mu^2)
=\tilde{f}_{q/q}^{(1)}({t},P_z,\tilde{\mu}^2)-f_{q/q}^{(1)}({t},\mu^2).
\label{eq:cqq1}
\end{eqnarray}
Both $\tilde{f}_{q/q}^{(1)}$ and ${f}_{q/q}^{(1)}$ can be calculated by using the Feynman diagrams in Fig.~\ref{fig:quasi-quark-lo}, but with $n_z\cdot A=0$ and $n\cdot A=0$ gauge, and $\gamma_z/2P_z$ and $\gamma^+/2P^+$ ``cut-vertex'', respectively.  We only give the derivation of $\tilde{f}_{q/q}^{(1)}$ here since the $f_{q/q}^{(1)}$ and its derivation are well known \cite{Collins:1981uw}.

By integrating out energy component of the gluon's momentum, we derive a compact expression,
\begin{eqnarray}
\tilde{f}_{q/q}^{(1)}(\tilde{x}, P_z,\tilde{\mu}^2)
=C_F \frac{\alpha_s}{2\pi} \frac{(4\pi)^\epsilon}{\Gamma(1-\epsilon)}
\int_0^{\tilde{\mu}^2} \frac{d{l_\perp^2}}{l_\perp^{2+2\epsilon}}
\int_{-\infty}^{+\infty} \frac{dl_z}{P_z}
\nnu \left[ \delta\left(1-\tilde{x}-y\right) - \delta\left(1-\tilde{x}\right) \right] \Bigg\{ \frac{1}{y}
\left(1-y + \frac{1-\epsilon}{2}y^2\right) \nnu \times\left[\frac{y}{\sqrt{\lambda^2+y^2}}+\frac{1-y}{\sqrt{\lambda^2+(1-y)^2}}\right]   + \frac{(1-y)\lambda^2}{2y^2 \sqrt{\lambda^2+y^2}} \nnu
  + \frac{\lambda^2}{2 y \sqrt{\lambda^2+(1-y)^2}}
+ \frac{1-\epsilon}{2} \frac{(1-y)\lambda^2}{\left[\lambda^2+(1-y)^2\right]^{3/2}} \Bigg\},
\label{eq:ft1}
\end{eqnarray}
where $y={l_z}/{P_z}$, $\lambda^2=l_\perp^2/P_z^2$,  $C_F=(N_c^2-1)/(2N_c)$ with $N_c=3$ is color factor, and the UV renormalization was imposed by a cutoff on $l_\perp^2$-integration.  In Eq.~(\ref{eq:ft1}), the $[\delta\left(1-\tilde{x}-y\right) - \delta\left(1-\tilde{x}\right)]$ dependence ensures the valence quark number conservation. The IR divergence from $1/y^2$ term, as $y\sim 0$, is naturally regularized by the $i\varepsilon$ prescription, and its net contribution is equal to taking the principle value of the $y$-integration.
The CO divergence, when $\lambda^2 \to 0$, is exactly the same as that of $f_{q/q}^{(1)}$.
Using Eq.~(\ref{eq:cqq1}) and $f_{q/q}^{(1)}$ in the $\overline{\rm MS}$ scheme, we obtain the NLO coefficient function,
\begin{eqnarray}\label{eq:C1}
\frac{{\cal C}_{q/q}^{(1)}({t})}{C_F \frac{\alpha_s}{2\pi}}
{\hskip -0.05in} &=& {\hskip -0.05in}
\left[\frac{1+{t}^2}{1-{t}} \ln\frac{\tilde{\mu}^2}{\mu^2} +1-{t}\right]_+
+ \Bigg[ \frac{t\Lambda_{1-{t}}}{(1-{t})^2}
+ \frac{\Lambda_{{t}}}{1-{t}}
\nnu
&\ & {\hskip -0.2in} +
\frac{ \text{Sgn}({t})\Lambda_{{t}}}{\Lambda_{{t}}+|{t}|}
- \frac{1+{t}^2}{1-{t}} \Big[ \text{Sgn}({t})\ln\left(1+\frac{\Lambda_{{t}}}{2|{t}|}\right)
\nnu
&\ & {\hskip -0.2in}
+ \text{Sgn}(1-{t}) \ln\left(1+\frac{\Lambda_{1-{t}}}{2|1-{t}|}\right) \Big]
 \Bigg]_N,
\label{eq:C1qq}
\end{eqnarray}
where $\Lambda_t=\sqrt{\tilde{\mu}^2/P_z^2 + t^2}-|t|$, $\text{Sgn}(t)=1$ if $t\ge 0$, and $-1$ otherwise.  In Eq.~(\ref{eq:C1qq}), the ``+"-function is conventional, and the ``$N$''-function is similarly defined as
\begin{eqnarray}
\int_{-\infty}^{+\infty} d{t} \Big[ g({t})\Big]_N h({t}) =  \int_{-\infty}^{+\infty} d{t}\, g({t}) \left[ h({t}) -h(1)\right],
\end{eqnarray}
where $h({t})$ is any well-behaved function.  When $P_z \to \infty$, $\Lambda_{t} = { O}(\tilde{\mu}^2/P_z^2)$, and terms within ``$[...]_N$'' in Eq.~\eqref{eq:C1qq} vanish, which is consistent with the large $P_z$ limit found in Ref.~\cite{Ji:2013dva}.  Our result for ${\cal C}_{q/q}^{(1)}(\tilde{x}/x)$ is consistent with the results obtained in Ref.~\cite{Xiong:2013bka}. As expected, the ${\cal C}_{q/q}^{(1)}$ in Eq.~(\ref{eq:C1qq}) is free of any IR and CO divergences, and in principle, depends on the choice of renormalization scheme for the quasi-PDFs.

\sect{Summary}
In summary, we proposed a QCD factorization based program to extract PDFs from LQCD calculations of good ``lattice cross sections", which on the one hand are LQCD calculable, and on the other hand are factorizable to PDFs. In this program, one first generates ``data'' from LQCD calculation of good LCSs, and then extracts PDFs by QCD global analysis of these ``data", similar to what have been done for extracting PDFs with experimental data. With today's computing power for LQCD calculation, our program for extracting PDFs is effectively doing ``low energy collision experiments" on LQCD, which is more relevant to PDFs at a relatively larger $x$, complementary to the global fitting program based on data from high energy scattering.

We took quasi-quark distribution as a case study to show that good LCSs could indeed exist. We demonstrated to all orders in pQCD that the multiplicatively renormalized quasi-quark distribution could be systematically factorized into PDFs with perturbative coefficient functions. That is, the renormalized quasi-quark distribution could serve as a good LCS, with a {\em finite} $P_z$ as ``collision energy'', and $\tilde{x} P_z$ as the hard scale defining the CO factorization.  We verified this explicitly by calculating the NLO coefficient function ${\cal C}^{(1)}_{q/q}$'s for quasi-quark distribution.

The precision of extracted PDFs from LQCD calculations could be greatly improved if there are more good LCSs, which is true \cite{Ma:2017pxb}.  Our proposed new LQCD based global fitting program to extract PDFs could be naturally extended to the study of transverse momentum dependent PDFs (TMDs) and generalized PDFs (GPDs), and other quark-gluon correlations of various hadrons.

{\it Acknowledgments}---We thank T.~Ishikawa, X.D.~Ji, G.~Sterman, S.~Yoshida and H.~Zhang for helpful discussions.  This work was supported in part by the U. S. Department of Energy under contract Nos.~DE-AC02-98CH10886 and DE-AC05-06OR23177, within the framework of the TMD Topical Collaboration, and the National Science Foundation under grant No.~PHY-0969739 and PHY-1316617.

\input{paper.bbl}

\end{document}

%% file: paper.bbl
\providecommand{\href}[2]{#2}\begingroup\raggedright\endgroup